\documentclass[aps,prl,twocolumn]{revtex4-1}
\usepackage[utf8]{inputenc}
\usepackage[english]{babel}
\usepackage[T1]{fontenc}
\usepackage{lmodern}
\usepackage{graphicx}
\usepackage{amsmath}
\usepackage{amsfonts}
\usepackage{amssymb}
\usepackage{microtype}
\usepackage{braket}
\usepackage{subfigure}
\usepackage{ulem}
\usepackage{etoolbox}
\usepackage[breaklinks=true,colorlinks, citecolor=blue]{hyperref}

\newcommand{\bi}{\bibitem}

\newcommand{\ct}{\cite}

\newcommand{\beq}{\begin{eqnarray}}
\newcommand{\eeq}{\end{eqnarray}}
\newcommand{\be}{\begin{equation}}
\newcommand{\ee}{\end{equation}}

\begin{document}

\title{Observing Dynamical Quantum Phase Transitions through Quasilocal String Operators}
\author{Souvik Bandyopadhyay$^1$, Anatoli Polkovnikov$^2$ and Amit Dutta$^1$} 
\affiliation{$^1$ Department of Physics, Indian Institute of Technology Kanpur, Kanpur 208016, India}
\affiliation{$^2$ Department of Physics, Boston University, Boston, Massachusetts, USA}

\begin{abstract}
	
We analyze signatures of the dynamical quantum phase transitions in physical observables. In particular, we show that both the expectation value and various out of time order correlation functions of the finite length product or string operators develop cusp singularities 
following quench protocols, which become sharper and sharper as the string length increases. We illustrated our ideas analyzing both integrable and nonintegrable one-dimensional Ising models showing that these transitions are robust both to the details of the model and to the choice of the initial state.		
\end{abstract}
\maketitle

Understanding out-of-equilibrium dynamics of quantum many body systems is an exciting field of recent research both from theoretical and experimental viewpoints \ct{polkovnikov11,calabrese_exact11,dutta15,eisert15,alessio16,jstat}. In this regard,  dynamical and equilibrium quantum  phase transitions (DQPTs), manifested as real-time singularities in time-evolving integrable and nonintegrable quantum systems, are indeed an emerging and intriguing phenomena \ct{heyl13,damski20,titum19,dasgupta_otoc15,roy17}.
{To probe DQPTs, a quantum many body system is  prepared initially  in the ground state $|\psi(0)\rangle$~{of some Hamiltonian}. At time $t=0$ a parameter $\lambda$ of the Hamiltonian in suddenly changed to say $\lambda=\lambda_f$ and  the subsequent  temporal evolution of the system generated by the time-independent final Hamiltonian $H(\lambda_f)$ is tracked.}   DQPTs occur at those instants of time $t$ when the evolved state  $|\psi(t)\rangle= \exp{[-iH(\lambda_{f})t]}|\psi(0)\rangle$, becomes orthogonal to the initial state $|\psi(0)\rangle$, i.e., the so called Loschmidt overlap (LO), ${\cal L}(t)= \left|\langle \psi(0)|\psi(t)\rangle \right|^2 $ vanishes. At those critical instants, the so called  dynamical free energy density (or the rate function of the return probability) defined as ${\cal F}
=-(1/N) \log |{\cal L}(t)|$, $N$ being the system size, develops nonanalytic singularities~{(cusps in 1D systems)} in the thermodynamic limit \ct{heyl13}. 

Following the initial proposal \ct{heyl13}, there have been a plethora of studies investigating intricacies of DQPTs in several  integrable and nonintegrable, one-dimensional  (as well as two-dimensional) quantum systems occurring subsequent to both sudden~\ct{karrasch13,kriel14,andraschko14,canovi14,heyl14,vajna14,
vajna15,schmitt15,palami15,heyl15,budich15,huang16,divakaran16,puskarov16,zhang16,heyl16,zunkovic16,sei17,fogarty17,heyl18,bhattacharya1,bhattacharya2,mera17,halimeh17,homri17,dutta17,sedlmayr181,bhattacharjee18,kennes18,piroli18,heyl_otoc18,nicola20,jafarilangari20,gurarie19}, and smooth~\ct{pollmann10,sharma15,sharma16,niccolo20} ramping protocols.  {The notion of DQPT} has also been generalized for {  mixed} initial states \ct{bandyopadhyay17,heyl_mixed17,abeling16,sedlmayr182} and finally also in open quantum systems \ct{bandyopadhyay18}. Analogous to equilibrium phase transitions, it has been established that one expects universal scaling of the dynamical free energy density near the critical instants with identifiable critical exponents. (For reviews on various aspects of DQPTs, we refer to Refs.~\ct{zvyagin17,victor17,heyl_review_18}.) 
Remarkably, these nonanalyticities have been detected experimentally subsequent to a rapid quench from a topologically trivial system into a Haldane-like system \ct{flaschner}.
 
Recently, DQPTs were also experimentally \ct{jurcevic16} detected in trapped ion setups simulating a long range interacting transverse field Ising model (TFIM). Starting from a degenerate ground state manifold, it was established that following a quench in an interacting chain of $^{40}{\rm Ca}^{+}$ ions, the dynamical free energy density develops cusp-like singularities at critical insants signalling DQPTs. However, a thorough understanding of the phenomena in harmony with the now well understood notion of equilibrium quantum phase transitions is far from being complete. {Although DQPTs may be characterized by a topological dynamical order parameter \ct{budich15,bhattacharya1} indicating the emergence of momentum space vortices at critical instants, there is an ongoing search for spatially local observables which are able to capture these nonequilibrium quantum phase transitions \ct{roy17,dasgupta_otoc15,asmi20}}. There has been many attempts aiming at finding real-time observable effects of the dynamical transitions on many body observables such as work distributions and the growth of entanglement in quenched systems \ct{nicola20,jafari20}. {A very interesting perspective on DQPTs was put forward in another recent trapped ion quantum simulator \ct{monroe17}, where the authors experimentally studied singularities in the domain wall statistics following a sudden quench of the transverse magnetic field in the system with long range Ising type interactions. }
\\

{One obvious drawback of DQPTs is that they are manifested in the overlap of the wave functions, which is difficult to observe experimentally. However, the experiment of J. Zhang {  et al.} (Ref.~\ct{monroe17}) showed that DQPTs can be also manifested in the behavior of nonlocal stringlike observables. But the precise mathematical connection between DQPTs and this experiment remains unclear.}  Currently, the broad questions which remain under scrutiny are: (i) Can the singular transitions at DQPTs be captured in the real-time behavior of strong observable quantities? (ii) What can one infer about the spatiotemporal locality of the observables required to detect the DQPTs? (iii) Is it possible to obtain the critical exponents associated with the dynamical transitions through a measurement of time-evolving observables \ct{heyl18}? In this work, we approach these issues by defining finite length string operators. {As the length of these operators increases they effectively play the role of the projection operators to polarized spin states.} Using an exact diagonalization scheme \ct{weinberg17,weinberg19}, we show that these observables are able to capture the critical singularities in their time evolving expectations and temporal correlators. We observe that early time rate function of the out of time order correlator (OTOC), which is an important quantity to study scrambling of information in chaotic systems \ct{swingle19} and quantum phase transitions \ct{heyl_otoc18,ceren19}, {quickly becomes} nonanalytic at the critical instants~{with the operator length} and show universal critical scaling near DQPTs. \\

To exemplify, we consider a ferromagnetic  transverse field Ising model (TFIM) with nearest and next nearest neighbor interactions ($J>0$ and $J_2>0$) and a noncommuting external field $h$, having $N$ spins,
\begin{equation}\label{eq:ham}
H=-J\sum\limits_{i=1}^{N}\sigma_z^i\sigma_z^{i+1}-J_2\sum\limits_{i=1}^{N}\sigma_z^i\sigma_z^{i+2}-h\sum\limits_{i=1}^{N}\sigma_x^i
\end{equation}
under periodic boundary conditions where $\sigma$'s are the Pauli Matrices satisfying standard ${\rm su}(2)$ commutation relations. The presence of both the nearest neighbour  and the next nearest neighbour interactions renders  the model nonintegrable with an integrable point at $J_2=0$. For~{concreteness} we start from a fully spin-polarized state $\ket{\uparrow\uparrow\uparrow...}$, {which is a ground state corresponding to} zero transverse field.  {However, this assumption can be lifted without affecting the results of this work as we will observe similar nonanalyticities in the infinite temperature OTOCs (also see Supplemental Material Ref.~\onlinecite{SM} for a discussion on generic initial states)}. As an observable we consider the translationally invariant Pauli string operator,
\begin{equation}\label{eq:project}
P_n(0)=\frac{1}{N}\sum\limits_{i=1}^{N}\frac{1}{2^n}\prod\limits_{i}^{i+n}\left(\mathbb{I}_i+\sigma_z^i\right),
\end{equation}
having a finite string size $n$ of a system of size $N$ and probe its time evolution following a sudden quench of the transverse field $h$ at $t=0$. When the strings in Eq.~\eqref{eq:project} span the whole system (i.e., when $n=N$), the quantity $P_n(0)$ simply reduces to the projector over the complete initial state $\ket{\psi(0)}$.

\begin{figure}[ht]
\includegraphics[width=1.0\columnwidth,height=6.0cm]{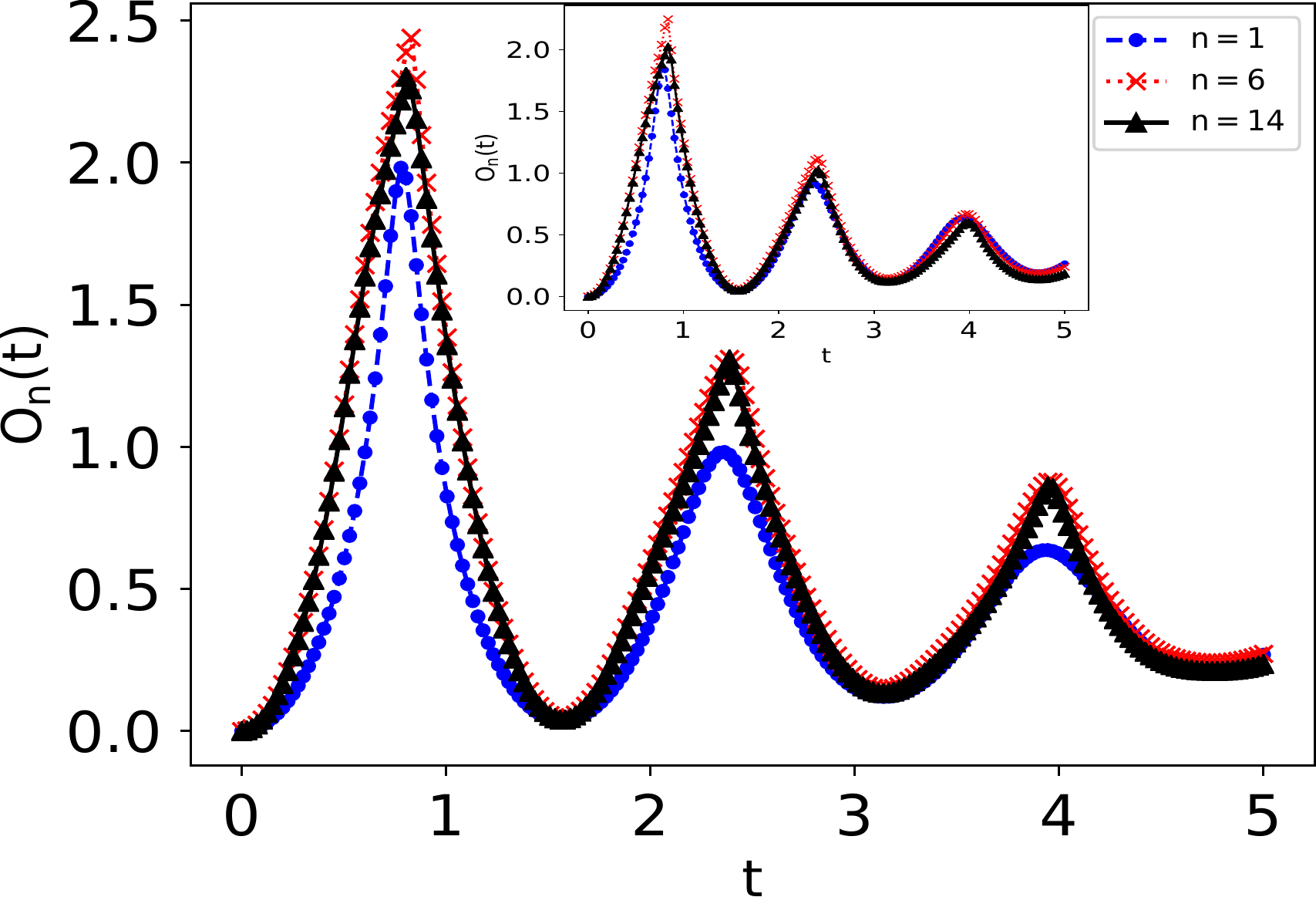}
\caption[]{(Color online) {Emergent cusp singularities in the observable $\mathcal{O}_n(t)$ (see Eq.~\eqref{eq:order1}) for finite string lengths $n$ following an integrable sudden quench in the transverse field (with $4J=1.0, 2J_2=0.0$ ), from the completely polarized ferromagnetic ground state ($2h=0$) to a paramagnetic phase ($2h=4.0$).  The emergence of the singularities are shown for various string lengths comparing the cusps with increasing string length $n$. (Inset) The same observable following a quench in the nonintegrable Ising chain ($4J=1.0, 4J_2=0.5$). The simulations have been performed in a chain containing $N=16$ spins using exact diagonalization.}}\label{fig:1}
\end{figure}

{Originally DQPTs are defined through emergent nonanalyticities of the rate function~\cite{heyl13,karrasch13} defined as
\begin{equation}
f(t)=-\lim\limits_{N\rightarrow\infty}\frac{1}{N}\log \mathcal{L}(t),
~\text{where}~{\mathcal{L}}(t)=\left|\braket{\psi(0)|\psi(t)}\right|^2.
\end{equation}
These nonanalyticities develop at critical instants of time ($t=t_c$).  It is easy to see that formally the rate function can be understood through the time-dependent expectation value of the ground state projection operator $P=|\psi(0)\rangle\langle \psi(0)|$ in the Heisenberg representation: $P(t)=\exp[i H t] P(0) \exp[-i H t]$, such that,
\be
f(t)=-\lim\limits_{N\rightarrow\infty}\frac{1}{N} \log  \braket{P(t)},
\ee
{where,} $\braket{\dots}=\braket{\psi(0)| \dots |\psi(0)}$.
In the one-dimensional Ising models, close to the critical point $\left|f(t)-f(t_c)\right|\sim |t-t_c|^\alpha$ with the universal critical exponent $\alpha=1$~\cite{heyl13, karrasch13, heyl15}.
For the initially polarized state $\ket{\uparrow\uparrow\uparrow...}$, clearly $P\equiv P_N(0)$. The idea of this work is to look into the expectation value of the operator $P_n(t)$ ($n\leq N$) instead of the full projector through the observable
\begin{equation}\label{eq:order1}
\mathcal{O}_n(t)=-\frac{1}{n}\log\braket{P_n(t)},
\end{equation}
We find (see Fig.~\ref{fig:1}) that the observable $\mathcal{O}_n(t)$ develops emergent cusp singularities at the critical instants in both integrable and integrability-broken systems, thus establishing that  DQPTs can be detected using normal physical observables. As it is evident from the plot the singularities quickly develop with increasing $n$, becoming very sharp for $n\gtrsim 6$ for the parameters used to generate this plot. The singularities were also seen to develop with increasing string length in the thermodynamic limit ($N\rightarrow\infty$) following an exact calculation for the integrable situation  {as the expectation $\braket{P_n(t_c)}$ vanishes at the critical instants exponentially fast in $n$ }\ct{SM}.
}

{These results can be generalized for quenches starting from an arbitrary initial state \ct{SM}, like a ground state corresponding to a finite transverse field within the paramagnetic phase or a mixed initial density matrix, for example, corresponding to a finite temperature ensemble. To see how it works, let us observe that for any initial 
\begin{figure*}
\subfigure[]{
\includegraphics[width=\columnwidth,height=6.5cm]{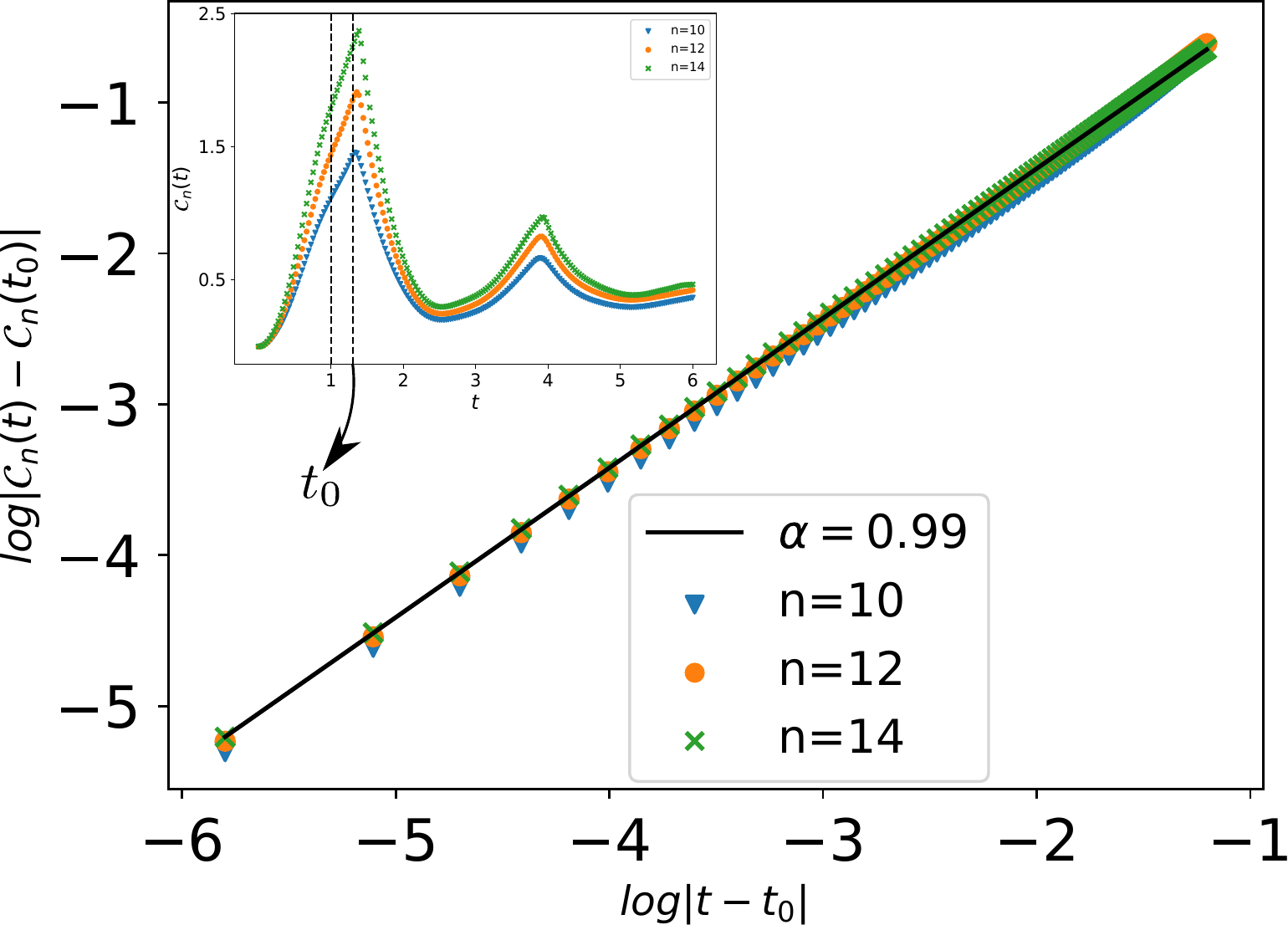}
\label{fig:2a}}
\subfigure[]{
\includegraphics[width=\columnwidth,height=6.5cm]{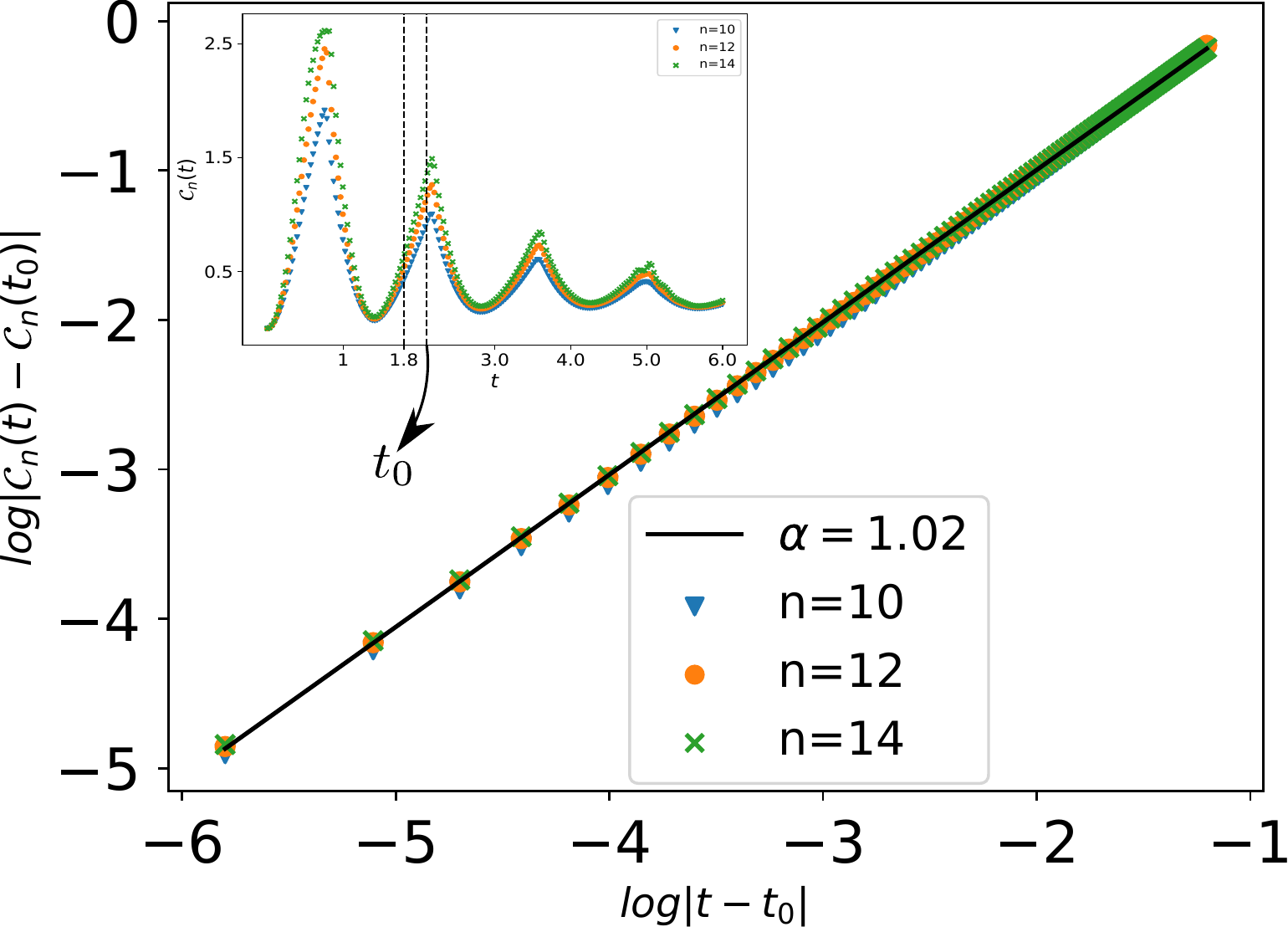}
\label{fig:2b}}
\caption{ (Color online) 
	{(a) The universal scaling of the OTOCs for an integrable chain  near the critical point. The deviation $\log|\mathcal{C}_n(t)-\mathcal{C}_n(t_0)|$ vs $\log|t-t_0|$ (in the region shown within the vertical dashed lines in inset) exhibits a scaling collapse for different string lengths $n$ (see also Ref.~\ct{halimeh20}); here we have chosen an instant $t_0=1.3$ to be a point near the critical point such that it captures the universal linear collapse region. (Inset) Corresponding emergent cusp singularities in the rate function of the OTOC $\mathcal{C}_n$ [see Eq.~\eqref{eq:oto}] for finite string lengths following a sudden quench in the transverse field (with $4J=1.5, 4J_2=0$ for $N=16$) from the completely polarized ferromagnetic ground state ($2h=0$) to a paramagnetic phase ($2h=2.5$). (b) The linear scaling of $\mathcal{C}_n$ (in the region shown within the vertical dashed lines in inset) and the corresponding cusps (Inset) at critical times in a nonintegrable chain (with $4J=1.0, 4J_2=1.0$ for $N=16$) subsequent to a quench from $2h=0.0$ to $2h=4.5$ and choosing a $t_0=2.1$ near the critical point. Both the integrable and the non-integrable situations exhibit a universal critical exponent of $\alpha\sim1$ as seen by the linear fits depicted by black solid lines.
	The linear scaling in both (a) and (b) has been shown for $t<t_c$ (left of the critical instant). Similar scaling has also been checked to hold for $t>t_c$.}}
\end{figure*}	
density matrix $\rho(0)$ we have,
\be
\label{eq:echo_OTOC}
C(t)\equiv{\rm Tr} [\rho(t) P(0) P(-t) P(0)]=p(t)\mathcal{L}(t),
\ee}
{where, $p(t) =\braket{\psi(0)|\rho(t) |\psi(0)}$. It is easy to check that, $${\rm Tr} [\rho(t) P(0) P(-t) P(0)]={\rm Tr} [\rho(0) P(t) P(0) P(t)],$$ such that, 
\be
C(t)=p(t)\mathcal{L}(t)= {\rm Tr} [\rho(0) P(t) P(0) P(t)].
\ee
The function $p(t)$ is a projection of the time-dependent density matrix to the
ground state $|\psi(0)\rangle$. In some situations, like when the initial state is the ground state of the initial Hamiltonian, $p(t)$ $=\mathcal{L}(t)$ such that $C(t)\propto \mathcal L^2(t)$. If the initial state, on the other hand, is stationary with respect to the final Hamiltonian $H(\lambda_f)$ then $p(t)={\rm const}(t)$ and $C(t)\propto \mathcal L(t)$. In both cases $-{1\over N}
\log[C(t)]$ is expected to show the exact same singularities as the rate function and scale similar to $f(t)$ with the same exponent $\alpha$ near the critical point.

The function $C(t)$ is nothing but the OTOC of the projection operator, which can be measured through, for example, quantum echo protocols~\cite{schmitt15,swan19,anatoli20,fine14,fine15}. {Like before instead of the full $C(t)$ we define the quasilocal truncated OTOC
\begin{equation}\label{eq:oto}
\mathcal{C}_n(t)=-\frac{1}{n}\log\braket{P_n(t)P_n(0)P_n(t)}, 
\end{equation}
where the average now is over the initial density matrix $\rho(0)$, which we first take to be the same as before $\rho(0)=|\psi(0)\rangle\langle \psi(0)|$ and later show that the results qualitatively do not change if we start from general mixed ensembles.
}

\begin{figure}[ht]
\centering
\includegraphics[width=0.9\columnwidth,height=5.75cm]{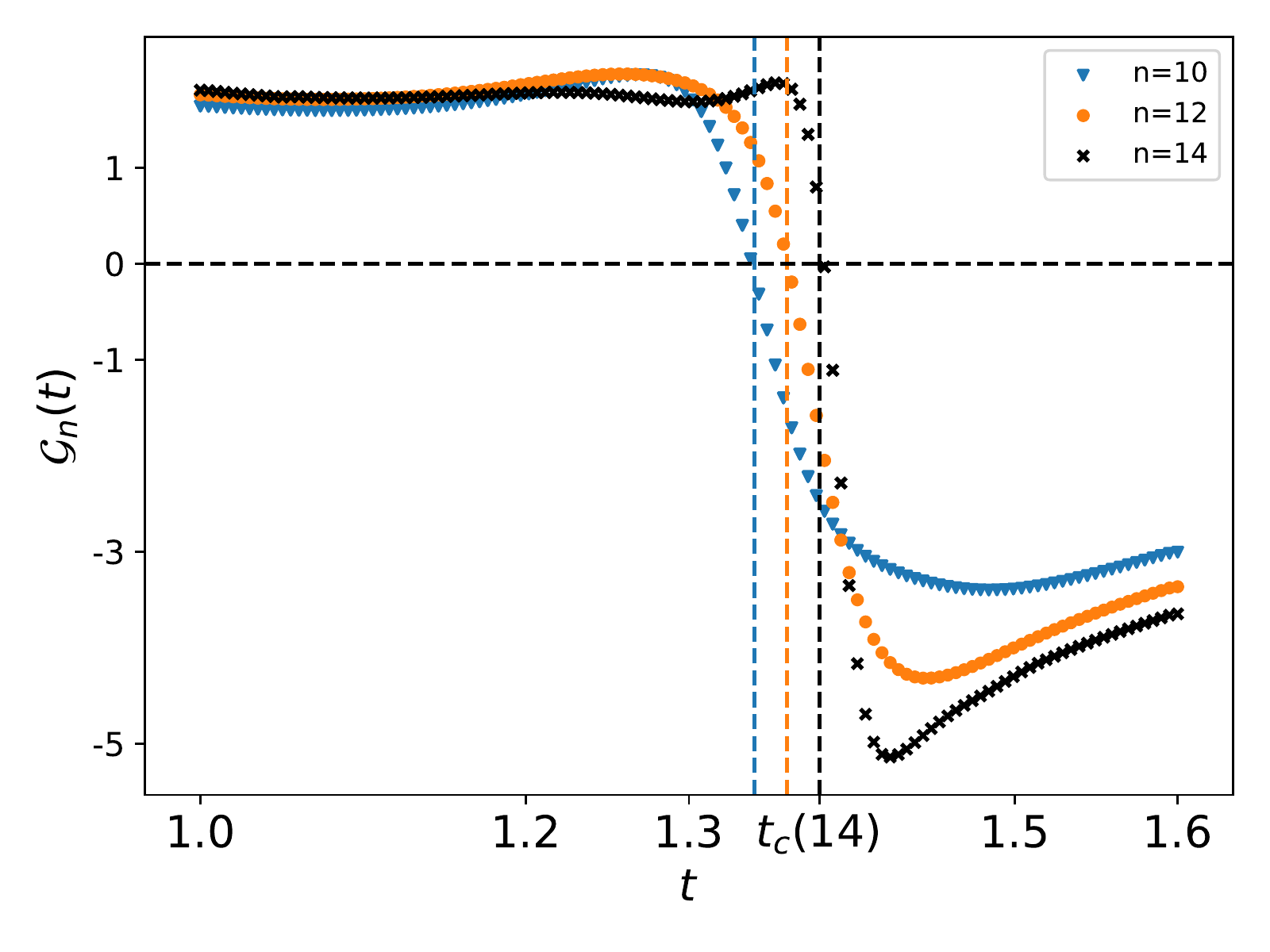}

\caption{ (Color online) 
	{Emergent jump discontinuities in the OTOC growth rate $\mathcal{G}_n$ [see Eq.~\eqref{eq:growth}] following an integrable sudden quench in the transverse field with the quench parameters being the same as in Fig.~\ref{fig:2a}.  The quantity $\mathcal{G}_n(t)$ vs $t$, is shown for various $n$ comparing the sharp jumps at the critical instant $t=t_c(n)$ with increasing string length. The critical instants $t_c(n)$ corresponding to the peaks in the rate function of the OTOC, for a string length $n$ can be determined by the emergent jump singularities in the OTOC growth rate (shown by vertical dashed lines in the figure where the rate crosses zero, with $t_c(14)$ explicitly marked). Similar jump discontinuities were also observed in the nonintegrable situations.}}
\label{fig:3} 
\end{figure}

Similar to the expectation $\mathcal{O}_n(t)$ the postquench rate function $\mathcal{C}_n(t)$ of the OTOC develops nonanalytic cusp singularities (see Fig.~\ref{fig:2a},\ref{fig:2b}) even for finite length string operators $P_n(0)$. We find that the observable $\mathcal{C}_n(t)$ apart from being singular at critical times, also shows a collapse to an universal scaling for sufficiently long strings near the critical point $t=t_c$, having critical exponent $\alpha\sim 1$ for quenches in both the integrable and nonintegrable chains. Consequently, the growth rate of the OTOC in its early time dynamics, shows singular discontinuous jumps at the critical instants of DQPTs. In Fig.~\ref{fig:3} we show that the jump singularities in the early time OTOC growth rate at critical instants,
\begin{equation}\label{eq:growth}
\mathcal{G}_n(t)=\frac{d\mathcal{C}_n(t)}{dt},
\end{equation}
for finite string operators $P_n(0)$ emerge with increasing sharpness for increasing string lengths.\\

Although we have demonstrated the development of emergent singularities through a quenched TFIM, the phenomena is not explicitly model dependent as has been previously established in literature. We also stress that the information about the  complete initial ground state is not a necessary requirement to study the nonequilibrium phase transitions which can also be detected in temporal correlators of the strings when summed over {  arbitrary} complete basis states. To elaborate, consider the infinite temperature autocorrelator and OTOCs,
\begin{eqnarray}\label{eq:tr_otoc}
\tilde{\mathcal{O}}_n(t)&=&-\frac{1}{n}\log{\rm Tr}\left[P_n(t)P_n(0)\right], \nonumber\\
\tilde{\mathcal{C}}_n(t)&=&-\frac{1}{n}\log{\rm Tr}\left[P_n(t)P_n(0)P_n(t)\right],
\end{eqnarray}
respectively. Remarkably, we find that though the traced correlations are independent of the full initial state $\ket{\psi(0)}$, they develop cusplike singularities at the critical instants for finite but sufficiently long strings and are therefore sufficient to observe the nonequilibrium transitions (see Fig.~\ref{fig:4}). This also establishes that the emergent critical behavior stays robust in the observables even when the dynamics starts from {  arbitrary} excited initial states.\\ 
\begin{figure}[ht]
	\centering
	\includegraphics[width=0.9\columnwidth,height=5.75cm]{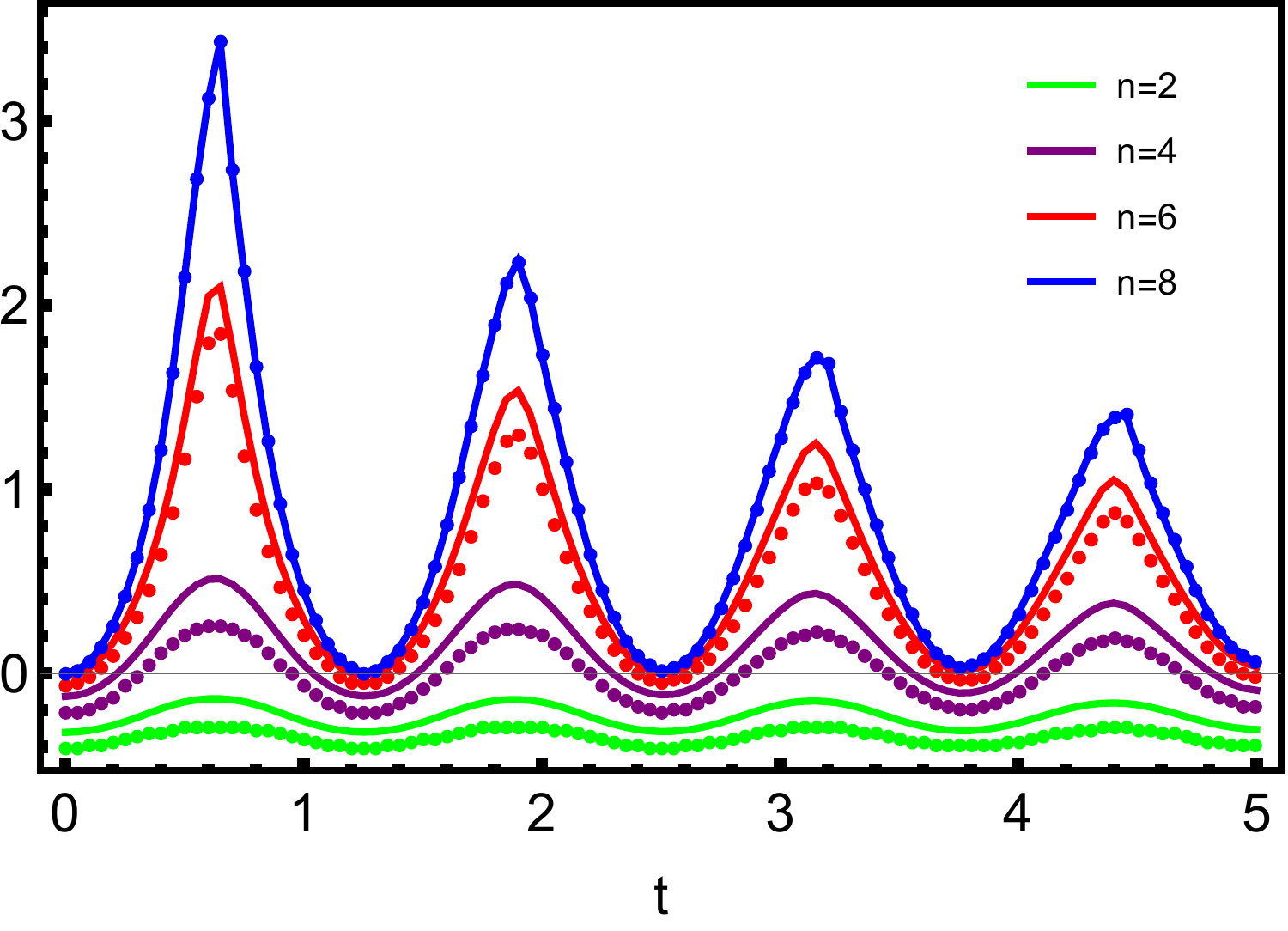}
	
	\caption{  {(Color online) 
			(Solid lines) The logarithm of the infinite temperature OTOC $\tilde{\mathcal{C}}_n(t)$ (see Eq.~\eqref{eq:tr_otoc}) at the critical instants show sharper singularities with increasing peak heights as the string size $n$ increases.} (dotted lines) The corresponding postquench autocorrelation functions $\tilde{\mathcal{O}}_n(t)$. The system is initially chosen to be in a completely polarized ground state of an integrable TFIM having a transverse field $h=0.0$, $J=0.1$ and suddenly quenched to a final field in the paramagnetic phase $h=2.5$ at $t=0$. The simulations have been performed for a chain of $N=8$ spins while considering finite strings of lengths $n$. The singularities were seen to become sharper for higher string sizes.}
	\label{fig:4} 
\end{figure}

{In conclusion, we demonstrated that DQPTs can be observed experimentally as postquench singularities developing in time for stringlike observables. These singularities become sharper with increasing string length. We showed that for the initial ground state these singularities emerge in the expectation values of string operators, while for generic, even infinite temperature, initial states they appear in the OTOC of such operators and can be detected through echo-type protocols. It is interesting that similar signatures of postquench singularities in two-spin observables were found in Refs.~\cite{dasgupta_otoc15,asmi20}. The precise relation of the results of that work to the present one regarding DQPTs is yet to be understood.}\\
 {One can also check (see Ref.~\ct{SM}) that the coefficient of variation of the observable $P_n(t_c)$ at a critical point remains finite and nonzero for local strings (having a finite length $n$) even in the thermodynamic limit, unlike that of the full projector $P_N(t_c)$ which diverges exponentially in system size. This suggests a further experimental advantage of the proposed observables as unlike the full Loschmidt overlap, it takes only a finite number of measurements to accurately determine the expectations of finite projectors even for infinitely large system sizes.}\\

\begin{acknowledgements}
S.B. acknowledges support from a PMRF fellowship, MHRD, India. A.D.  and A.P. acknowledge financial support from SPARC program, MHRD, India. A.P. was supported by NSF DMR-1813499 and AFOSR FA9550-16-1-0334.
We acknowledge Sourav Bhattacharjee and Somnath Maity for comments. We acknowledge QuSpin and HPC-2010, IIT Kanpur for computational facilities.
\end{acknowledgements}

\begin{flushleft}
	{  Note added:}\end{flushleft}  {During preparation of the manuscript we came across a similar study \ct{halimeh20} which also exhibit the efficacy of local operators in capturing DQPTs.}\\

\newpage
\setcounter{equation}{0}
\setcounter{figure}{0}
\setcounter{table}{0}
\makeatletter
\renewcommand{\theequation}{S\arabic{equation}}
\renewcommand{\thefigure}{S\arabic{figure}}
\renewcommand{\citenumfont}[1]{S#1}
\renewcommand{\bibnumfmt}[1]{[S#1]}
\renewcommand{\labelenumi}{(\theenumi)}
\renewcommand{\theenumi}{\roman{enumi}}
\renewcommand{\bibnumfmt}[1]{[S#1]}

\widetext

\setcounter{page}{1}
\begin{center}
	\textbf{\large Supplemental material to "Observing Dynamical Quantum Phase Transitions through Quasilocal String Operators"}
\end{center}
	
	\section{An analytical approach to the integrable chain}
	\label{sec:long}
	Here we consider the integrable transverse field Ising model in a canonically rotated frame under periodic boundary conditions,
	\begin{equation}\label{eqs:ham}
		H=-J\sum\limits_{j}\sigma^x_j\sigma^x_{j+1}+h\sum\limits_{j}\sigma^z_j,
	\end{equation}
	where the transverse field is acting in the $z$ direction and,
	\begin{equation}
		\sigma_j^i\equiv\sigma_{(j~mod~N)}^i,
	\end{equation}
	$N$ being the total number of spins, taken here to be even. It is well known that the transverse Ising chain is integrable in the sense that it can be mapped to a free fermionic system using the Jordan Wigner transformation,
	\begin{eqnarray}
		\sigma^z_j&=&1-2c^{\dagger}_{j}c_j,\nonumber\\
		\sigma^x_i&=&\left( c_i+c_i^{\dagger}\right)\prod\limits_{j<i}\left(1-2c_j^{\dagger} c_j\right),
	\end{eqnarray}
	where $c_j^{\dagger}$ and $c_j$ are fermionic creation and annihilation operators satisfying standard anti-commutation relations. The transformed fermionic chain takes the following BCS form,
	\begin{equation}
		H=-J\sum\limits_{j}\left(c_j^{\dagger}-c_j\right)\left(c_{j+1}^{\dagger}+c_{j+1}\right)+2h\sum\limits_{j}c_j^{\dagger}c_j.
	\end{equation}
	Depending of the fermion parity specified by the operator,
	\begin{eqnarray}
		P_{\pm}&=&\frac{1}{2}\left(\mathbb{I}\pm\Sigma\right), ~\text{where,}\nonumber\\
		\Sigma&=&\prod\limits_{j}\left(1-2c_j^{\dagger}c_j\right),
	\end{eqnarray}
	it can be shown that the Hamiltonian can be decoupled into two independent parity sectors,
	\begin{equation}
		H=P_{+}HP_{+}+P_{-}HP_{-}.
	\end{equation}
	Further, under periodic boundary conditions, the chain is translationally invariant which allows the construction of definite momentum decoupled subspaces following a Fourier transform, described by the Hamiltonian,
	\begin{equation}
		H=\bigoplus_{k>0}H(k)=\bigoplus_{k}\begin{pmatrix}
			c_k^{\dagger} & c_{-k}
		\end{pmatrix}\begin{pmatrix}
			2(h-J\cos{k}) & -2J\sin{k} \\
			-2J\sin{k} & 2(J\sin{k}-h) 
		\end{pmatrix}\begin{pmatrix}
			c_k \\ c_{-k}^{\dagger},
		\end{pmatrix}
	\end{equation} 
	such that,
	\begin{equation}\label{eqs:transform}
		c_j=\frac{1}{\sqrt{N}}\sum\limits_{-\pi<q<\pi}e^{iqj}c_q.
	\end{equation} 
	In the thermodynamic limit of infinitely many spins ($N\rightarrow\infty$), the modes $k$ span a continuous set of momenta. We choose the system to initially be in the BCS ground state of the Ising Hamiltonian belonging in the even parity sector for the parameters $h=h_i=0$ and $J=J_i>0$,
	\begin{equation}
		\ket{\psi(0)}=\bigotimes_{k}\ket{\psi_k(0)}=\bigotimes_{k}\left(u_k(0)-v_k(0) c_k^{\dagger}c_{-k}^{\dagger}\right)\ket{0},
	\end{equation}
	where $\ket{0}$ is the fermionic vacuum. Starting from this state, at time $t=0$, a sudden quench is inflicted upon the transverse field to $h_i\rightarrow h$ and the subsequent evolution of the system is tracked. The time dependent state is then given by,
	\begin{equation}\label{eqs:psit}
		\ket{\psi(t)}=\bigotimes_{k}\ket{\psi_k(t)}=\bigotimes_{k}\left(u_k(t)-v_k(t) c_k^{\dagger}c_{-k}^{\dagger}\right)\ket{0},
	\end{equation}
	where the quantities $u_k(t)$ and $v_k(t)$ satisfy Bogoluibov system of equations,
	\begin{equation}
		i\frac{d}{dt} \begin{pmatrix}
			v_k(t) \\ u_k(t)
		\end{pmatrix}=\begin{pmatrix}
			2(h-J\cos{k}) & -2J\sin{k} \\
			-2J\sin{k} & 2(J\sin{k}-h) 
		\end{pmatrix}\begin{pmatrix}
			v_k(t) \\ u_{k}(t)
		\end{pmatrix}.
	\end{equation}
	As elaborated in the main text, we evaluate the expectation of the operators $P_n$ over the time evolved state $\ket{\psi(t)}$, and subsequently measure the observable,
	\begin{equation}\label{eqs:P}
		\mathcal{O}_n(t)=-\frac{1}{n}\log{\left<P_n(t)\right>},
	\end{equation}
	for finite $n$, to observe the dynamical quantum phase transitions (DQPTs).  The full projector $P_N$ is such that it simply projects onto the completely polarized ferromagnetic state $\ket{\rightarrow\rightarrow\rightarrow...}$.  To exemplify,
	\begin{equation}
		P_n=\frac{1}{N}\sum\limits_{j}\prod\limits_{i=0}^{n-1}\left(\frac{\mathbb{ I}+\sigma^x_{j+i}}{2}\right),
	\end{equation}
	such that,
	\begin{eqnarray}
		P_2&=&\frac{1}{2^2N}\sum\limits_{j}\left(\mathbb{ I}+\sigma_j^x+\sigma^x_{j+1}+\sigma_j^x\sigma^x_{j+1}\right),\\
		P_3&=&\frac{1}{2^3N}\sum\limits_{j}\left(\mathbb{I}+\sigma_j^x+\sigma_{j+1}^x+\sigma^x_{j+2}+\sigma_j^x\sigma^x_{j+1}+\sigma_j^x\sigma^x_{j+2}+\sigma_{j+1}^x\sigma^x_{j+2}+\sigma_j^x\sigma^x_{j+1}\sigma_{j+2}^x\right),
	\end{eqnarray}
	and so on. To evaluate the quantity in Eq.~\eqref{eqs:P}, we cast the operator $P_n$ in terms of the Jordan Wigner fermionic operators and take their expectations over the time evolved state $\ket{\psi(t)}$. The advantage of starting from a definite parity eigenstate $\ket{\psi(0)}$ is that since the Hamiltonian $H$ respects parity, the time evolved state $\ket{\psi(t)}$ does not change its parity in time. Therefore, the expectation over single spin operators such as $\braket{\psi(t)|\sigma^x_j|\psi(t)}$ identically vanish when the initial state lies in the even parity sector. Cosequently, the expectation $\braket{P_1}$ stays invariant in times. Similarly, $\braket{P_2}$ and $\braket{P_3}$ in its fermionic representation simply reduces to,
	\begin{eqnarray}
		\braket{P_2(t)}&=&\frac{1}{2^2N}\sum\limits_{j}1+X_1^j(t),\nonumber\\
		\braket{P_3(t)}&=&\frac{1}{2^3N}\sum\limits_{j}1+2X_1^j(t)+X_2^j(t),
	\end{eqnarray}
	where,
	\begin{eqnarray}
		X_1^j(t)&=&\braket{c_j^{\dagger}c_{j+1}^{\dagger}}+\braket{c_j^{\dagger}c_{j+1}}+{\rm cc},\nonumber\\
		X_2^j(t)&=&\braket{c_j^{\dagger}c_{j+2}^{\dagger}}+\braket{c_j^{\dagger}c_{j+2}}+\braket{c_j^{\dagger}c_{j+1}^{\dagger}c_{j+2}^{\dagger}c_{j+1}}
		-\braket{c_j^{\dagger}c_{j+1}^{\dagger}c_{j+1}c_{j+2}}+{\rm cc},
	\end{eqnarray}
	such that $cc$ represents the complex conjugate terms and $\braket{.}\equiv\braket{\psi(t)|.|\psi(t)}$. In a similar fashion, all higher length string expectations $\braket{P_n}$ for $n>3$ can also be expressed in terms of a finite series of time dependent correlation functions of fermionic operators,
	\begin{equation}
		X(t)=\sum\limits_{j}\sum\limits_{m}A_m\braket{a_j^ma_{j+1}^ma_{j+2}^m....},
	\end{equation}
	such that $a_j^m\in\{\mathbb{I}_j,c_j^{\dagger}$, $c_j$\} and coefficients $A_m\in\mathbb{C}$. However, owing to the noninteracting nature of the Hamiltonian, the time evolved state can be expressed in a gaussian form and the larger correlations can indeed be decomposed in terms of the two point fermionic correlations using Wick's theorem\cite{calabrese11} such as,
	\begin{equation}
		\braket{a_j^ma_{j+1}^ma_{j+2}^ma_{j+3}^m}=\braket{a_j^m a_{j+1}^m}\braket{a_{j+2}^ma_{j+3}^m}-\braket{a_j^ma_{j+2}^m}\braket{a_{j+1}^ma_{j+3}^m}+\braket{a_j^ma_{j+3}^m}\braket{a_{j+1}^ma_{j+2}^m}.
	\end{equation}
	To evaluate the two point correlations in the time dependent state $\ket{\psi(t)}$, an easier route is to evaluate momentum correlation functions and the two point correlation functions are then connected to them via a Fourier transform due to the translational invariance. The momentum correlations in our scenario takes the form,
	\begin{eqnarray}\label{kcorr}
		\braket{c_{k}^{\dagger}c_q}&=&\delta_{kq}\left|v_k(t)\right|^2,~\text{and}~\nonumber\\
		\braket{c_kc_q}&=&\delta_{q,-k}u_{k}^{*}(t)v_k(t),
	\end{eqnarray}
	where $u_k(t)$ and $v_k(t)$ are the Bogoluibov functions in the time evolved state $\ket{\psi(t)}$ in Eq.~\eqref{eqs:psit}. The corresponding real-space two point correlations are then obtained by the Fourier transform (see Eq.~\eqref{eqs:transform}),
	\begin{eqnarray}
		C(l,t)=\braket{c_{j+l}^{\dagger}c_{j}}&=&\frac{1}{\sqrt{N}}\sum\limits_{k}e^{ikl}\left|v_k(t)\right|^2\nonumber ~\text{and}~\\
		F(l,t)=\braket{c_{j+l}c_j}&=&\frac{1}{\sqrt{N}}\sum\limits_{k>0}2i\sin{(kl)}u_{k}^{*}(t)v_k(t).
	\end{eqnarray}
	Hence, all the string expectations $\braket{\psi(t)|P_n|\psi(t)}$ for any value of $n$ can be expressed in terms of the correlation functions $C(l,t)$ and $F(l,t)$ with varying lengths $l$ at all times. In Fig.~\ref{fig:S1}, we have shown the time evolution of the observables $\mathcal{O}_n$ for $n=2,3$ and $4$ in a thermodynamically large chain ($N\rightarrow\infty$). As reported in the main text, $\mathcal{O}_n(t)$ develops cusp singularities at critical instants following a sudden quench, thus capturing dynamical phase transitions effectively.
	\begin{figure*}[h]
		\centering
		\includegraphics[width=.57\textwidth,height=6cm]{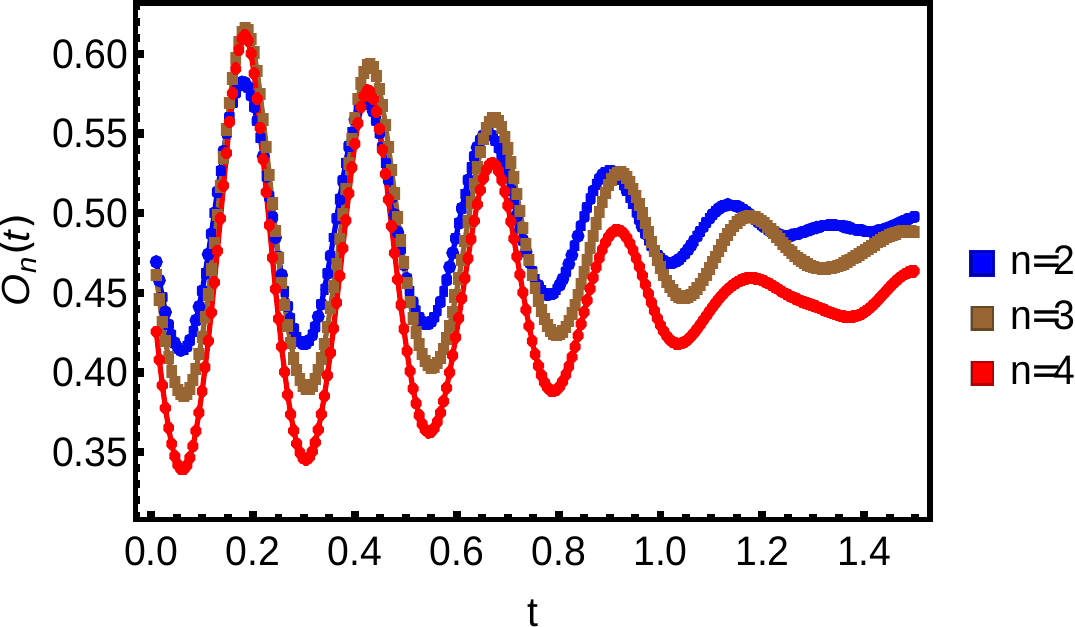}
		
		\caption{ (Color online) The post quench evolution of the observable $\mathcal{O}_n(t)$ (defined in Eq.~\eqref{eqs:P}) developing cusp singularities at critical instants for increasing string length in a thermodynamically large integrable Ising chain ($N\rightarrow\infty$) under periodic boundary conditions. The system was initially prepared in the ground state for the parameters $h=0$ and $J=0.75$. At $t=0$, the transverse magnetic field $h$ was suddenly quenched to $h=6.5$ and the system was left to evolve with the quenched Hamiltonian.}
		\label{fig:S1} 
	\end{figure*}
	
	\section{Projectors in the transverse direction}
	\label{smsec:transverse}
	
	In this section we exhibit the emergent development of cusps in the postquench dynamics while starting from deep into the paramagnetic phase and quenching to the ferromagnetic phase. Here in, we choose the initial state to be the completely polarized paramagnetic state $\ket{\psi(0)}=\ket{\uparrow\uparrow\uparrow...}$ as the eigenstate of the Ising model (Eq.~\eqref{eqs:ham}) with a large transverse magnetic field ($h_i\gg J_i$). At $t=0$, the system is quenched into the ferromagnetic phase ($J_f>h_f$) and the subsequent evolution is studied using the obervable,
	\begin{eqnarray}\label{eqs:oz}
		\mathcal{O}_n^z(t)=-\frac{1}{n}\log{\braket{P_n^z(t)}},
	\end{eqnarray}
	such that the operator,
	\begin{eqnarray}
		P_n^z(0)=\frac{1}{N}\sum\limits_{j}\prod\limits_{i=0}^{n-1}\left(\frac{\mathbb{ I}+\sigma^z_{j+i}}{2}\right),
	\end{eqnarray}
	is the finite version of the projector onto the transverse polarized state, i.e., $P_{N}^z(0)=\ket{\psi(0)}\bra{\psi(0)}$. Recasting in the fermionic language we obtain the string proportional to the transverse magnetisation ($n=1$) and the bilinear ($n=2$) spin projectors as,
	\begin{eqnarray}
		P_1^z(0)&=&\frac{1}{2N}\sum\limits_{j}Z_1^j,\nonumber\\
		P_2^z(0)&=&\frac{1}{2^2N}\sum\limits_{j}Z_2^j,
	\end{eqnarray} 
	where,
	\begin{eqnarray}
		Z_1^j=2-2c_j^{\dagger}c_j,~\text{and}\nonumber\\
		Z_2^j=4-8c_j^{\dagger}c_j+4c_j^{\dagger}c_jc_{j+1}^{\dagger}c_{j+1}.
	\end{eqnarray}
	We then follow a similar procedure as elaborated in Sec.~\ref{sec:long} to determine the time dependent expectations,
	\begin{eqnarray}\label{p1}
		\braket{P_n^z(t)}=\lim\limits_{N\rightarrow\infty}\frac{1}{2^nN}\sum\limits_{j}\braket{\psi(t)|Z_n^j|\psi(t)}.
	\end{eqnarray}
	As noted previously, all the higher order correlators can then be subsequently decomposed into strings of two point correlation functions using Wick's theorem and the observable $\mathcal{O}_n^z(t)$ can then be evaluated in a straight forward way. In principle this complete protocol can be done for any order $n$ of the observables.
	\begin{figure*}[h]
		\centering
		\includegraphics[width=0.65\textwidth,height=6.5cm]{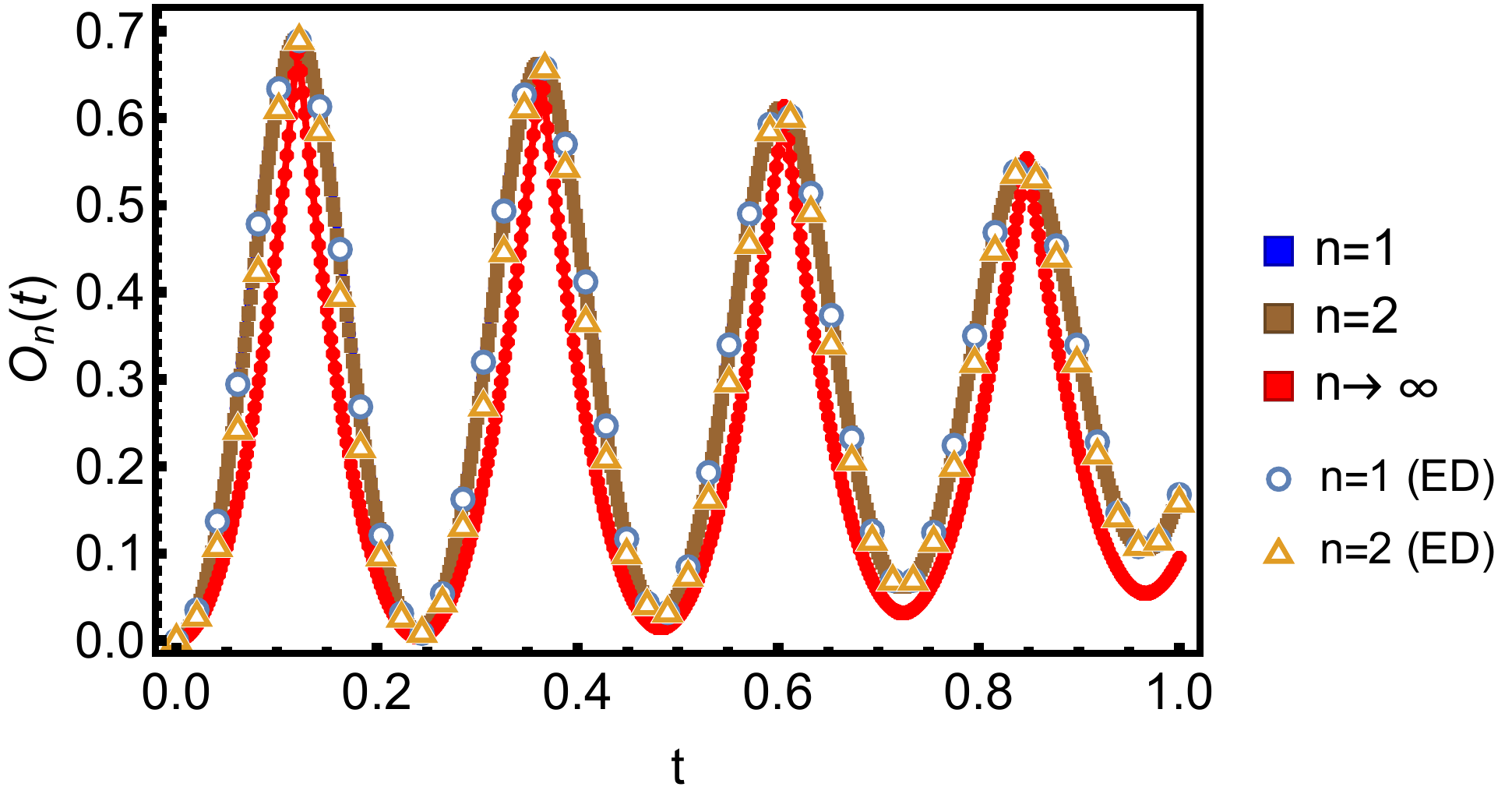}
		
		\caption{ (Color online) The post quench evolution of the observable $\mathcal{O}_n^z(t)$ (defined in Eq.~\eqref{eqs:oz}) developing cusp singularities at critical instants for increasing string length in a thermodynamically large integrable Ising chain ($N\rightarrow\infty$) under periodic boundary conditions and a comparison with the numerical ED scheme for 16 spins. The system was initially prepared in the completely polarized paramagnetic ground state for the parameters $J=0$ and $h=0.5$. At $t=0$, the nearest neighbor interaction $J$ was suddenly quenched to $J=6.5$ while keeping the transverse magnetic field invariant. Subsequently the system was left to evolve with the quenched Hamiltonian. The complete dynamical free energy density $f(t)$ (see Eq.~\eqref{eqs:f}) is seen to develop cusp singularities exactly at the critical instants predicted by the quasi local observables $\mathcal{O}_n^z(t)$.}
		\label{fig:S2} 
	\end{figure*}
	
	In Fig.~\ref{fig:S2}, we demonstrate the development of emergent cusp singularities in the local observables $\mathcal{O}^z_n(t)$ following the quench considered from a paramagnetic polarized state to the ferromagnetic phase in a thermodynamically large chain ($N\rightarrow\infty$). We also observe the sharp nonanalyticities in the complete projector (for $n=N$), i.e. the rate function of the Loschmidt overlap itself,
	\begin{equation}\label{eqs:f}
		f(t)=\lim\limits_{N\rightarrow\infty}\mathcal{O}_N^z(t)=-\lim\limits_{N\rightarrow\infty}\frac{1}{N}\log\left|\braket{\psi(0)|\psi(t)}\right|^2=-\frac{1}{2\pi}\int_0^{\pi}dk\log\left|\braket{\psi_k(0)|\psi_k(t)}\right|^2,
	\end{equation}
	and conclude that the emergent singularities in the local observables $\mathcal{O}_n^Z(t)$ are indeed at the same instants where the complete rate function becomes singular, thus reflecting the dynamical criticality of DQPTs completely. We reiterate that the string observables in principle can be exactly calculated for an arbitrary order $n$ following a quench in the integrable system.

	\section{Scaling analysis of the observable at critical instants with system and string size}
	\begin{figure*}
		\centering
		\subfigure[]{
			\includegraphics[width=0.52\textwidth, height=6.2cm]{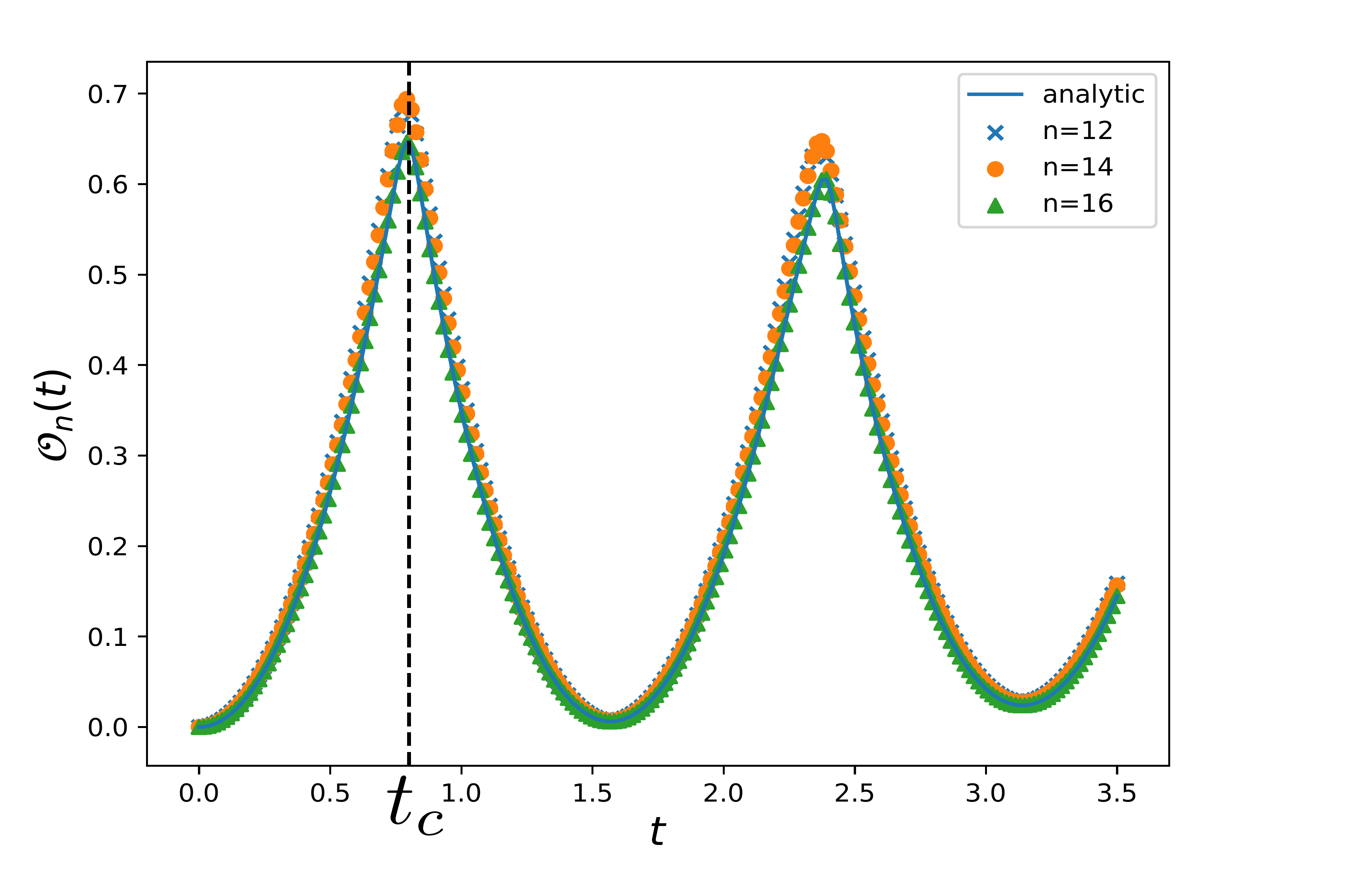}\label{var1}
		}	
		\subfigure[]{
			\includegraphics[width=0.45\textwidth, height=6cm]{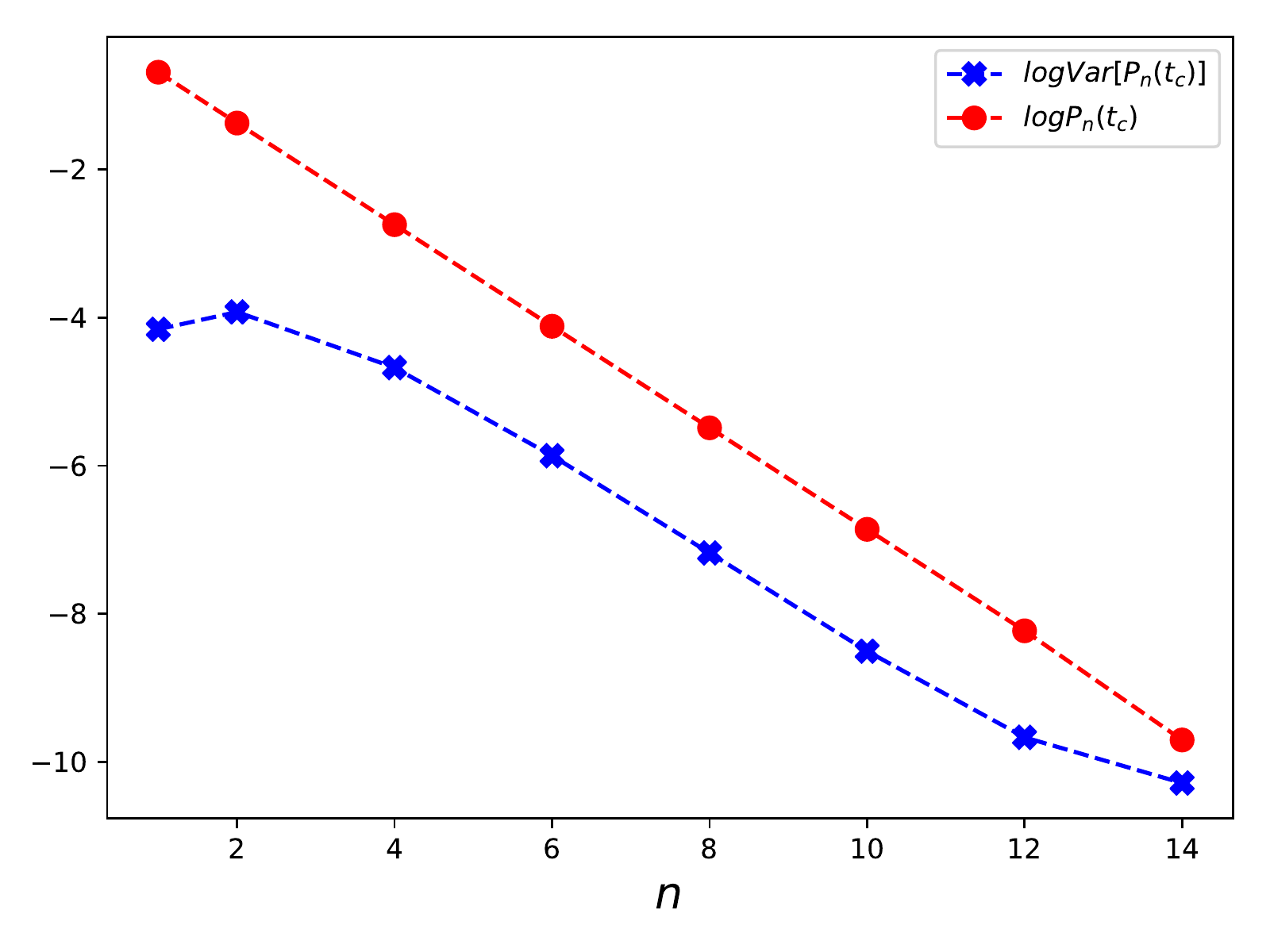}
			\label{var2}}
		\subfigure[]{
			\includegraphics[width=0.45\textwidth, height=6cm]{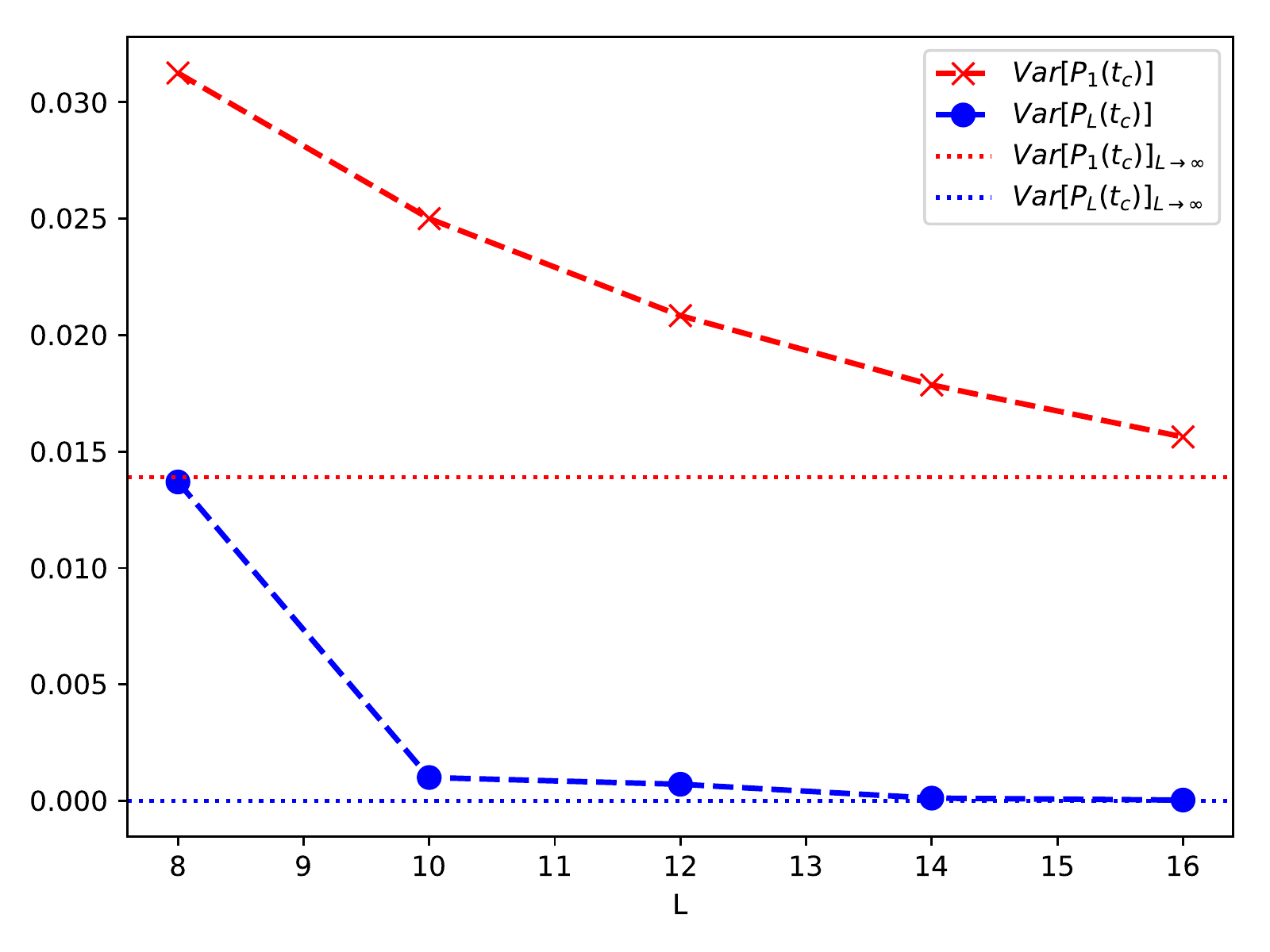}
			\label{var3}}	
		\hspace{1.2cm}\subfigure[]{
			\includegraphics[width=0.45\textwidth, height=6cm]{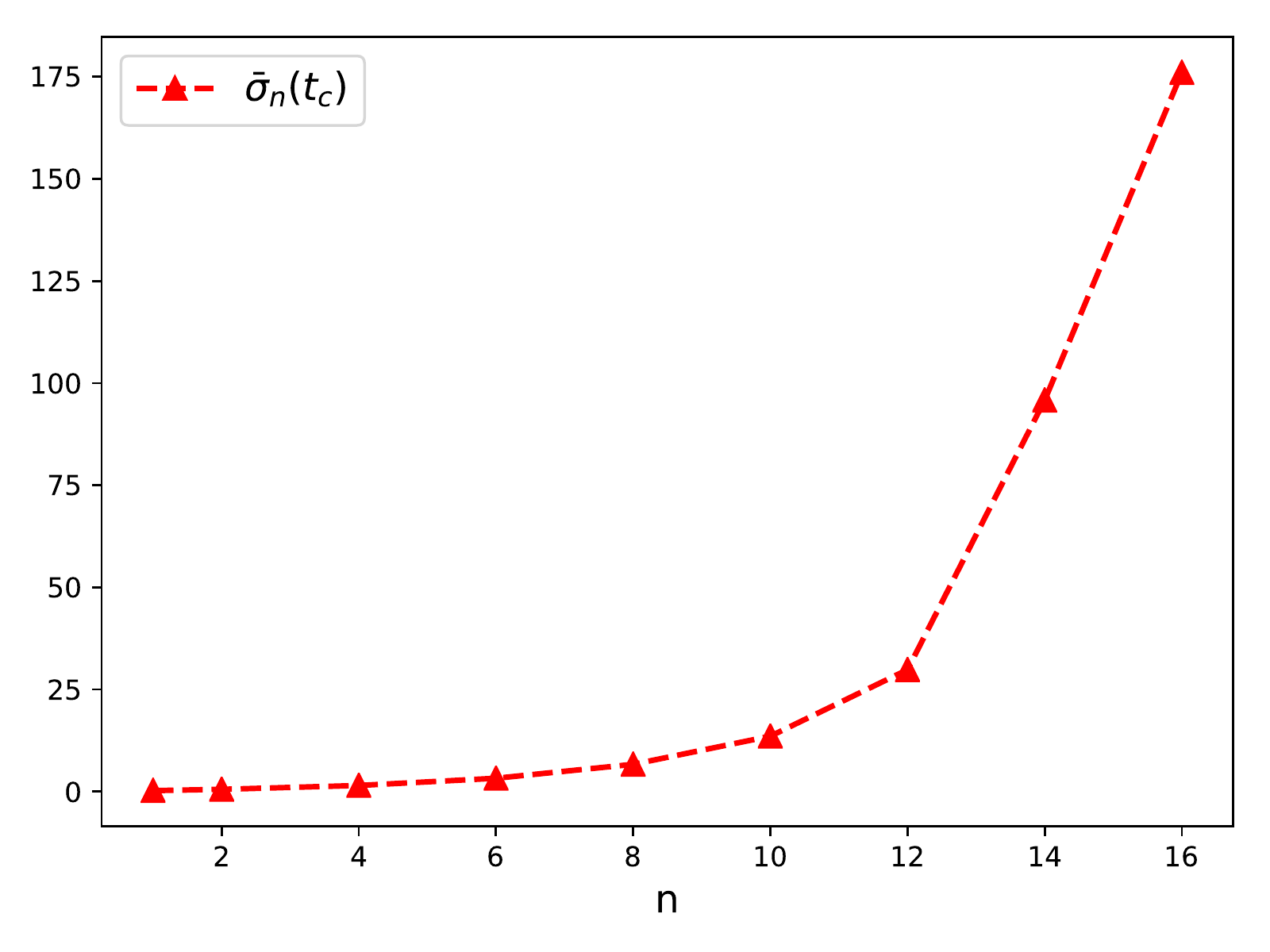}
			\label{var4}}
		\subfigure[]{
			\includegraphics[width=0.5\textwidth, height=6cm]{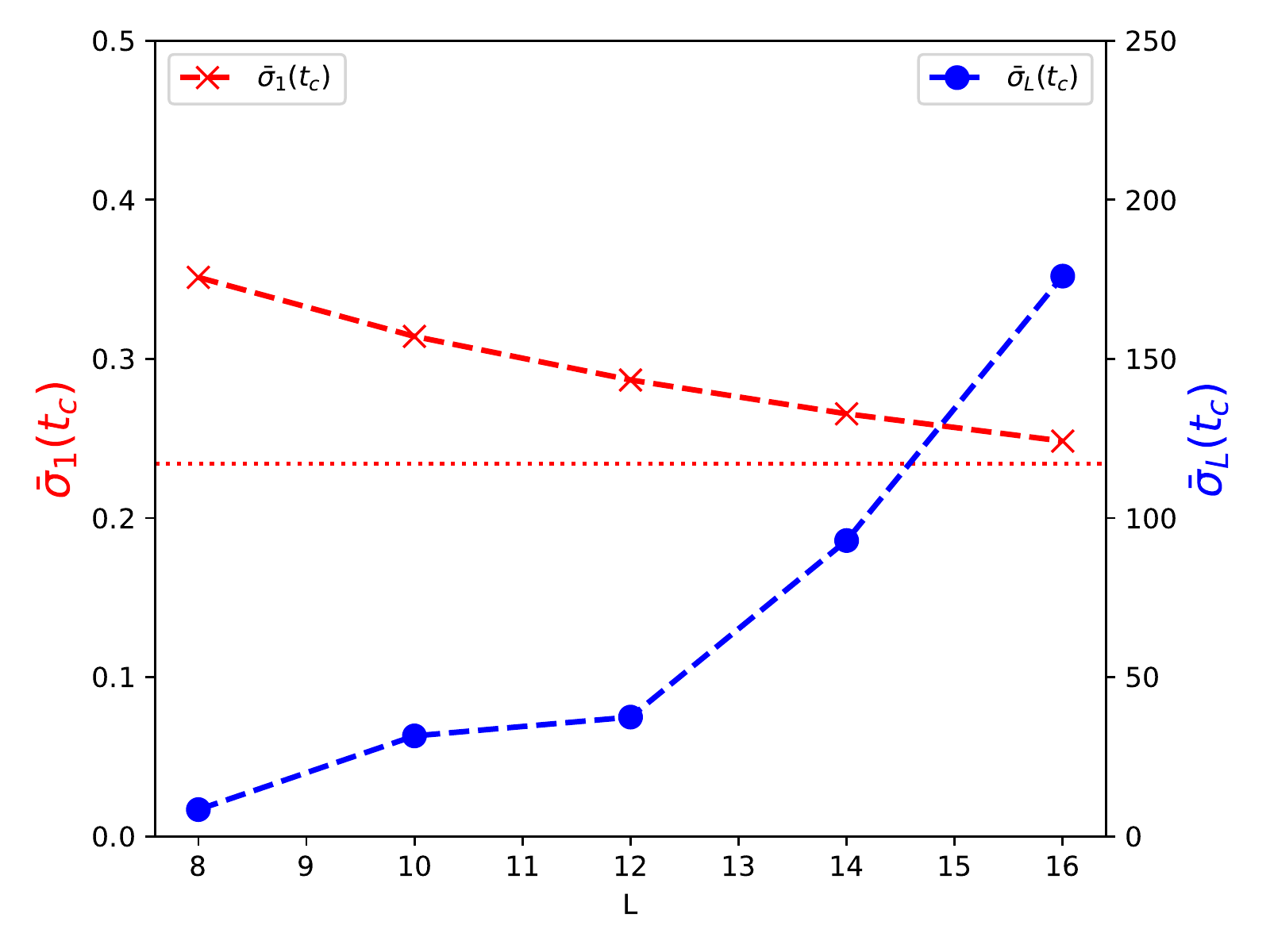}
			\label{var5}}
		
		\caption{(Color online) (a) Emergent cusp singularities in the observable $\mathcal{O}_n(t)$ for finite string lengths in a system size of $L=16$ spins following a quench from the paramagnetic phase $h=0.1$, $J=0.0$ to the ferromagnetic phase $h=0.1$ and $J=1.0$ in the integrable Ising model. The vertical dashed line shows a critical instant $t=t_c$. (b) The logarithm of the variance of $P_n(t_c)$ and the expectation $\braket{P_n(t_c)}$ with increasing string length $n$ for a fixed $L=16$. As seen in the figure, $\braket{P_n(t_c)}$ approaches zero with increasing string size exponentially fast in $n$. Also, the variance of $P_n(t_c)$ for a fixed $L$ decreases as the string length $n$ increases. (c) The variance of the full projector $P_L(t_c)$ vs the local projector $P_1(t_c)$ showing that unlike the full projector, the finite projectors have nonzero variances at the critical point even in the thermodynamic limit. The analytical values of the corresponding quantities in the thermodynamic limit are shown by horizontal dotted lines. (d) The relative standard deviation of finite projectors $P_n(t_c)$ at the critical point increases with string size for a system size of $16$ spins. (e) A comparision between the relarive standard deviation of the full projector and that of the finite projector $P_1$ at the critical instant shows that $\tilde{\sigma}_L(t_c)$ diverges with increasing system size while $\tilde{\sigma}_1(t_c)$ approaches a finite value with increasing system size. The dotted horizontal line shows the analytical value for $\tilde{\sigma}_1(t_c)$ in the thermodynamic limit.}
	\end{figure*}
	In this section we analyse the collapse of the expectation $\braket{P_n(t)}$ and its variance ${\rm Var}[P_n(t)]$ to zero (which in turn manifests as the cusp singularities in $\mathcal{O}_n(t)$) as one increases the string length $n$  for a system size $L$. Starting from a completely polarized paramagnetic ground state $\ket{\psi(0)}=\ket{\uparrow\uparrow\uparrow...}$, we subject the system to a sudden quench at time $t=0$ to the ferromagnetic phase while keeping the system integrable throughout. As discussed previously, subsequent, to the quench, the observable $\mathcal{O}_n(t)$ develop nonanalyticities at critical instants $t_c$ (see for example Fig.~S\ref{var1}), where $\mathcal{O}_n(t)$ is given by Eq.~\eqref{eqs:oz}. When the string size spans the complete system, i.e., for $n=L$, $P_n$ essentially is the projector onto the initial ground state $\ket{\psi(0)}$. It is therefore expected that the expectation $\braket{P_L(t)}$ being the Loschmidt overlap must vanish at the critical instants in a thermodynamically large system ($L\rightarrow\infty$).\\

	However, in finite size systems, the expectation $\braket{P_L(t_c)}$ or equivalently the Loschmidt overlap at the critical point approaches zero exponentially in the system size. This can be seen from Fig.~S\ref{var2} where we show the variance,
	\begin{equation}\label{eq:var1}
		{\rm Var}\left[P_n(t)\right]=\braket{P_n(t)^2}-\braket{P_n(t)}^2,
	\end{equation}
	and mean $\braket{P_n(t)}$, with respect to the string length $n$ at the critical point $t=t_c$ (shown in Fig.~S\ref{var1}). As is evident from Fig.~S\ref{var2}, the expectation $\braket{P_n(t_c)}$ approaches zero exponentially in the string size $n$ for a fized system size $L$. This also justifies the fact that the expectation of the full projector $\braket{P_L(t_c)}$ falls off to zero exponentially with the system size. This resonates with the fact that analogous to equilibrium quantum phase transitions, true nonanalyticities in the dynamical free energy density $\mathcal{O}_L(t_c)$ emerge only in the thermodynamic limit of the system.\\
	
	The variance of the finite projectors $P_n(t_c)$ for a fixed system size $L$ approaches zero as $n$ increases but nevertheless remains finite for finie string lengths $n$. This finiteness of the variance remains preserved for finite strings even in the thermodynamic limit of the system. To show this, using Eq.~\eqref{p1} and Eq.~\eqref{eq:var1} we analytically calculate the variance of $P_n(t_c)$ for a fixed $n=1$ in a thermodynamically large system, 
	\begin{equation}
		{\rm Var}\left[P_1(t_c)\right]=\frac{1}{L^2}\sum\limits_{ij}1-\braket{c_i^{\dagger}c_i}-\braket{c_j^{\dagger}c_j}+\braket{c_i^{\dagger}c_ic_j^{\dagger}c_j}-\left(\frac{1}{L}\sum\limits_l1-\braket{c_l^{\dagger}c_l}\right)^2.
	\end{equation}
	Utilizing the translation invariance of the system and Wick's decomposition, the variance can be expressed in terms of the momentum space correlators and hence is straightforward to evaluate in infinitely large systems. Fig.~S\ref{var3} shows that the variance of $\braket{P_1(t_c)}$ indeed approaches the finite nonzero value obtained analytically for an infinitely large system, as the system size increases.\\

	On the contrary, if one considers the variance of the full projector $P_L(t_c)$, one obtains,
	\begin{equation}\label{eq:sfp}
		{\rm Var}\left[P_L(t_c)\right]=\braket{P_L(t_c)^2}-\braket{P_L(t_c)}^2=\braket{P_L(t_c)}-\braket{P_L(t_c)}^2.
	\end{equation}
	Since, the expectation of the full projector $P_L(t_c)$ is equivalent to the Loschmidt overlap itself, it must vanish at critical instants in the thermodynamic limit (as also seen in Fig.~S\ref{var2}). This shows that the variance of the complete projector at critical times vanishes exponentially in system size whereas for finite strings, the variance remain nonzero even in the thermodynamic limit. However, the coefficient of variation/relative standard deviation $\tilde{\sigma}_{L}(t_c)$ of the full projector $P_L(t_c)$ at the critical point diverges exponentially fast in system size,
	\begin{equation}
		\lim\limits_{L\rightarrow\infty}\tilde{\sigma}_{L}(t_c)=\lim\limits_{L\rightarrow\infty}\frac{\sqrt{{\rm Var}[P_L(t_c)]}}{\braket{P_L(t_c)}}=\lim\limits_{L\rightarrow\infty}\frac{\sqrt{1-\braket{P_L(t_c)}}}{\sqrt{\braket{P_L(t_c)}}}\rightarrow\infty,
	\end{equation}
	where the second equality follows from Eq.~\eqref{eq:sfp}. Therefore, to accurately measure the complete projector near the critical points in a thermodynamically large system, one needs to perform exponentially large number of measurements. On the other hand as we show in Fig.~S\ref{var4}, for finite strings, the relative standard deviation $\tilde{\sigma}_n(t_c)$ remains finite at the critical instants. In Fig.~S\ref{var5} we demonstrate that $\tilde{\sigma}_n(t_c)$ for finite string lengths remains finite in the thermodynamic limit while $\tilde{\sigma}_L(t_c)$ diverges in system size.\\

	This in turn suggests that unlike the complete Loschmidt overlap, it requires only a finite number of measurements to accurately determine the expectation of projectors having a finite string length $n$ even in a thermodynamically large system.
	
	\section{Generality of the choice of the initial state}
	
	In the manuscript, we have demonstrated all the ideas using a completely spin polarized initial state to retain the simplicity of the discussion. However, as we have argued in the main manuscript, the results hold for arbitrary initial states as we explicily show in this section. Considering the initial state $\ket{\psi(0)}$ to be a ground state of the system with a finite nonzero transverse magnetic field and also a nearest neighbor Ising coupling, it is straight forward to see that following a quench generated by the Hamiltonian given in Eq.~(1) of the manuscript with $J_2=0$,
	\begin{equation}
		\braket{P_L(t)}=\left|\braket{\psi(t)|\uparrow\uparrow\uparrow..}\right|^2,
	\end{equation}
	where $\ket{\uparrow\uparrow\uparrow..}$ is again the completely spin  polarized state, when the operator $P_n(0)$ is given by Eq.~(2) of the manuscript. Therefore, the observable $\mathcal{O}_L(t)$ develops emergent cusp singularities whenever the time dependent state $\ket{\psi(t)}$ becomes orthogonal to the state $\ket{\uparrow\uparrow\uparrow..}$. In Fig.~S\ref{init0} and Fig.~S\ref{init1}, we show that these emergent nonanalyticities are also captured by the quasilocal observable $\mathcal{O}_n(t)$ when $n$ is finite even when the initial state is not completely polarized. However, it must be noted that since the initial state for such a generic scenario is not $\ket{\uparrow\uparrow\uparrow..}$ onto which $P_L(0)$ projects, the critical instants of the DQPTs exhibited by $\mathcal{O}_n(t)$ might differ from those in the case of the completely polarized initial state.
	
	\begin{figure*}
		\centering
		\subfigure[]{
			\includegraphics[width=0.6\textwidth, height=7.5cm]{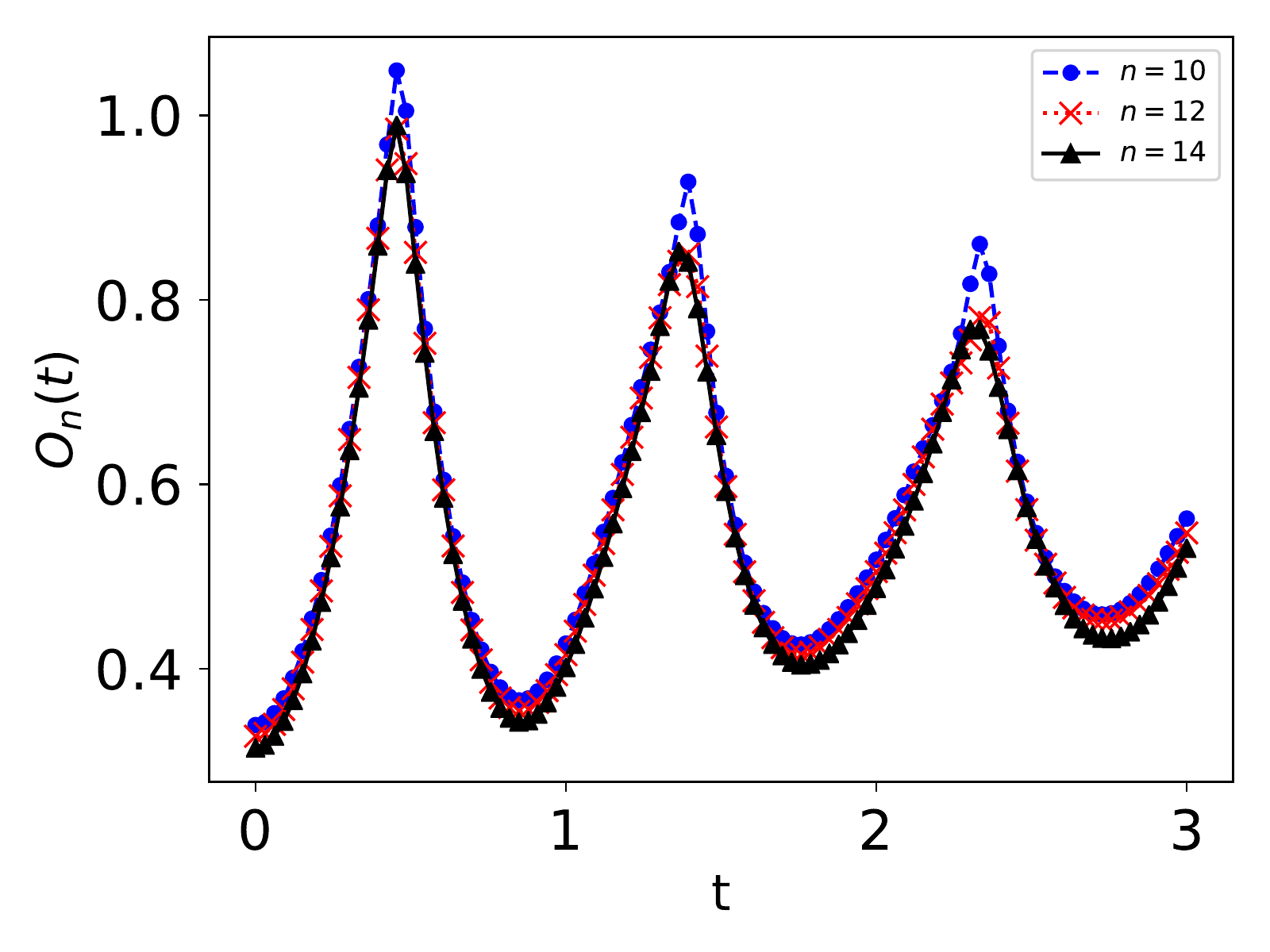}
			\label{init0}}	
		\subfigure[]{
			\includegraphics[width=0.6\textwidth, height=7.5cm]{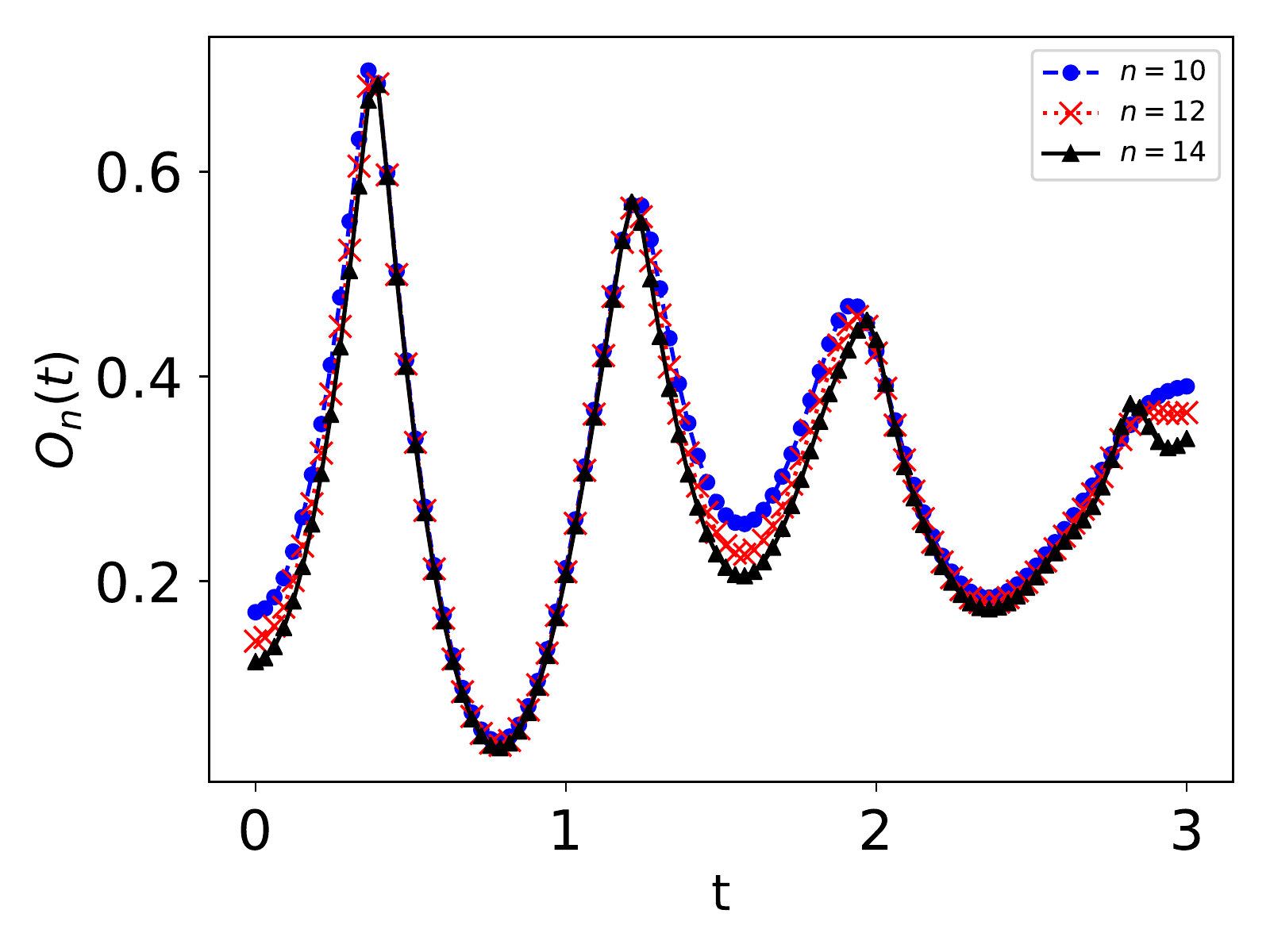}
			\label{init1}}
		
		\caption{(Color online) (a) Cusp singularities in $\mathcal{O}_n(t)$ (defined in Eq.~(5) of the main manuscript) following a sudden quench from the ground state of the integrable Ising Hamiltonian (Eq.~(1) of the main manuscript with $J_2=0$) having a finite transverse field $2h=1.0$ and $4J=1.5$ to $2h=4.0$ and $4J=1.5$. (b) The same observable following a quench from a completely polarized ground state of the system with $2h=0$ and $4J=1.5$ to $2h=4.0$ and $4J=1.5$. Both the simulations have been performed on a chain containing $16$ spins.}
	\end{figure*}

\end{document}